\documentclass[twocolumn]{aastex63}
\usepackage{amssymb,amsmath}
\usepackage{graphicx}
\usepackage{xcolor,natbib}
\usepackage{multirow}
\usepackage[T1]{fontenc}
\usepackage{ae,aecompl}
\usepackage{newtxtext, newtxmath}
\usepackage{float}
\usepackage{subfigure}
\usepackage{longtable}
\usepackage{enumitem}
\usepackage{longtable,listings}
\usepackage[flushleft]{threeparttable}
\usepackage{parcolumns}
\usepackage{hyperref}

\def\oc3{[O~{\sc iii}]$_c$}
\def\ob3{[O~{\sc iii}]$_b$}
\def\obj{SDSS J2308}
\def\dif{\mathop{}\!\mathrm{d}}

\submitjournal{ApJ}

\shorttitle{A high-redshift TDE candidate}
\shortauthors{Zhang XueGuang}
\begin{document}

\title{A central TDE candidate detected through spectroscopic continuum emission properties 
in a SDSS blue quasar}

\correspondingauthor{XueGuang Zhang}
\email{xgzhang@gxu.edu.cn}
\author{XueGuang Zhang$^{*}$}
\affiliation{Guangxi Key Laboratory for Relativistic Astrophysics, School of Physical 
Science and Technology, GuangXi University, Nanning, 530004, P. R. China}

\begin{abstract} 
	In this manuscript, properties of spectroscopic continuum emissions are considered 
to detect potential tidal disruption event (TDE) candidates among SDSS quasars. After 
considering the simple blackbody photosphere model applied to describe quasar continuum 
emissions with parameters of blackbody temperature $T_{BB}$ and blackbody radius $R_{BB}$, 
SDSS quasars and reported optical TDEs occupy distinct regions in the space of $T_{BB}$ 
and $R_{BB}$. Then, through the dependence of $R_{BB}$ on $T_{BB}$ for SDSS quasars, 402 
outliers in SDSS Stripe82 region can be collected. Among the 402 outliers, the \obj~ at 
$z=1.16$ is mainly considered, due to its SDSS spectrum observed around the peak brightness 
of the light curves. With the 7.2-year-long light curves described by theoretical TDE model, 
the determined $T_{BB}$ and $R_{BB}$ through its spectroscopic continuum emissions are 
consistent with the TDE model determined values, to support the central TDE. Moreover, 
considering simulated results on continuum emissions of SDSS quasars around $z\sim1.16$, 
confidence level higher than 4$\sigma$ can be confirmed that the continuum emissions of 
\obj~ are not related to normal quasars. Furthermore, accepted CAR process to simulate 
intrinsic AGN variability, the confidence level higher than $3\sigma$ can be confirmed 
that the long-term light curves of \obj~ are related to a central TDE. Jointed the 
probabilities through both spectroscopic and photometric simulations, the confidence level 
higher than $5\sigma$ can be confirmed to support the central TDE in \obj.
\end{abstract}


\section{Introduction}

	Quasars have two fundamental characteristics in their spectroscopic results, broad 
emission lines from central broad emission line regions (BLRs) and continuum emissions from 
central black hole (BH) accreting systems, as shown and discussed in \citet{sm00, sh11, oy15, 
zh21b, ir23}. Meanwhile, similar spectroscopic properties of broad emission lines from BLRs 
and continuum emissions from BH accreting systems can also be expected and detected in 
tidal disruption events (TDEs) with fallback debris from tidally disrupted stars accreted 
onto central BHs.

	Theoretical and/or observational continuum emissions related to TDEs can be found in 
\citet{re88, lu97, gm06, ve11, ce12, gr13, wz17, wy18, mg19, tc19, ry20, lo21, sg21, vg21, 
zl21, nl22, zs22, yr23, zh23a, ss24}. Meanwhile, among the reported optical TDE candidates, 
apparent broad Balmer and/or Helium emission lines related to TDEs debris are fundamental 
optical spectroscopic characteristics in some cases, such as the reported broad Balmer 
emission lines in TDE candidates in \citet{ve11, ht14, md15, ht16, lz17, ht19, sn20, hf20, 
hi21, zh21, zh24a, zh24b}, the reported broad Helium emission line in TDE candidates in 
\citet{gs12, ht16, ht16b, bl17, gr19}, and theoretical discussions on BLRs related to TDEs 
debris in \citet{gm14}, and the reported broad emission lines in the more recent small 
samples of TDEs in \citet{cl22, yr23}. Detailed review on theoretical TDEs can be found 
in \citet{st19}. And more recent review on observational properties of TDEs can be found 
in \citet{gs21}.

	Besides the similar spectroscopic features of broad emission lines and continuum 
emissions, however, there are some different intrinsic continuum emissions between quasars 
and TDEs. An observed optical spectrum at one given epoch from a TDE can be described by 
the accepted Blackbody photosphere model with two parameters of photosphere temperature 
$T_{BB}$ and photosphere radius $R_{BB}$, such as the detailed discussions in \citet{mg19}. 
Moreover, as shown in \citet{mg19} for TDEs, the TDEs expected $T_{BB}$, $R_{BB}$ and 
bolometric luminosity are around $10^{4-5}$K, $10^{14-15}$cm and $10^{44-45}{\rm erg/s}$ 
at the epoch around the peak brightness of TDEs expected long-term light curves. Due to 
different time evolution properties of TDEs with main-sequence stars with different stellar 
masses disrupted by SMBHs with different BH masses, properties of $T_{BB}$ and $R_{BB}$ 
are mainly considered on the epoch around the TDEs expected peak brightness in this manuscript.

	However, for quasars, such as the SDSS quasars discussed in \citet{gh05} and in 
\citet{sh11}, the corresponding bolometric luminosities could be around $10^{45}{\rm erg/s}$ 
or higher, after considering the bolometric corrections (around 10-15) applied to the 
continuum luminosities at rest wavelength 5100\AA~ as discussed in \citet{rg06, db20, nh20, 
sf24}. Therefore, quasars should have higher bolometric luminosities (and higher optical 
continuum luminosities) than TDEs. Meanwhile, as will shown in the next section, the 
continuum emissions in pointed wavelength bands in quasars can also be simply described 
by the Blackbody photosphere model which has been applied in TDEs, leading to different 
estimated parameters of $T_{BB}$ and $R_{BB}$ between quasars and TDEs. To compare 
properties of $T_{BB}$ and $R_{BB}$ between quasars and TDEs should provide further clues 
to detect potential TDE candidates which have probably similar spectroscopic features as 
those of quasars, which is the main objective of this manuscript.

\begin{figure*}
\centering\includegraphics[width = 18cm,height=6cm]{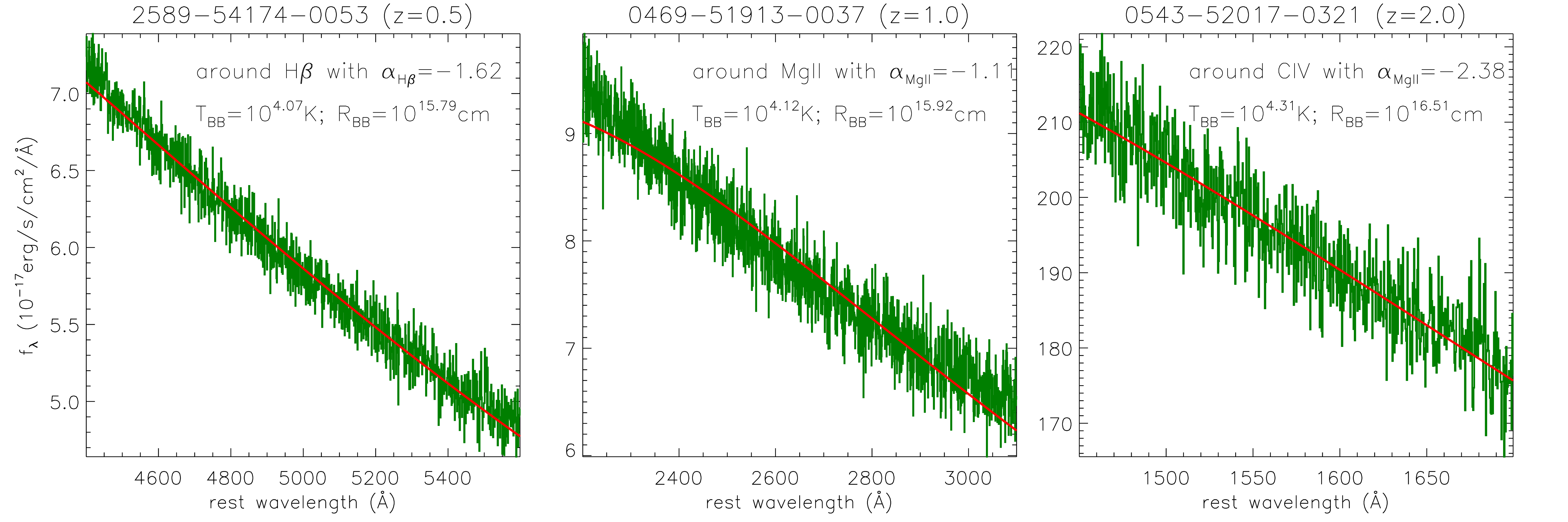}
\centering\includegraphics[width = 18cm,height=6cm]{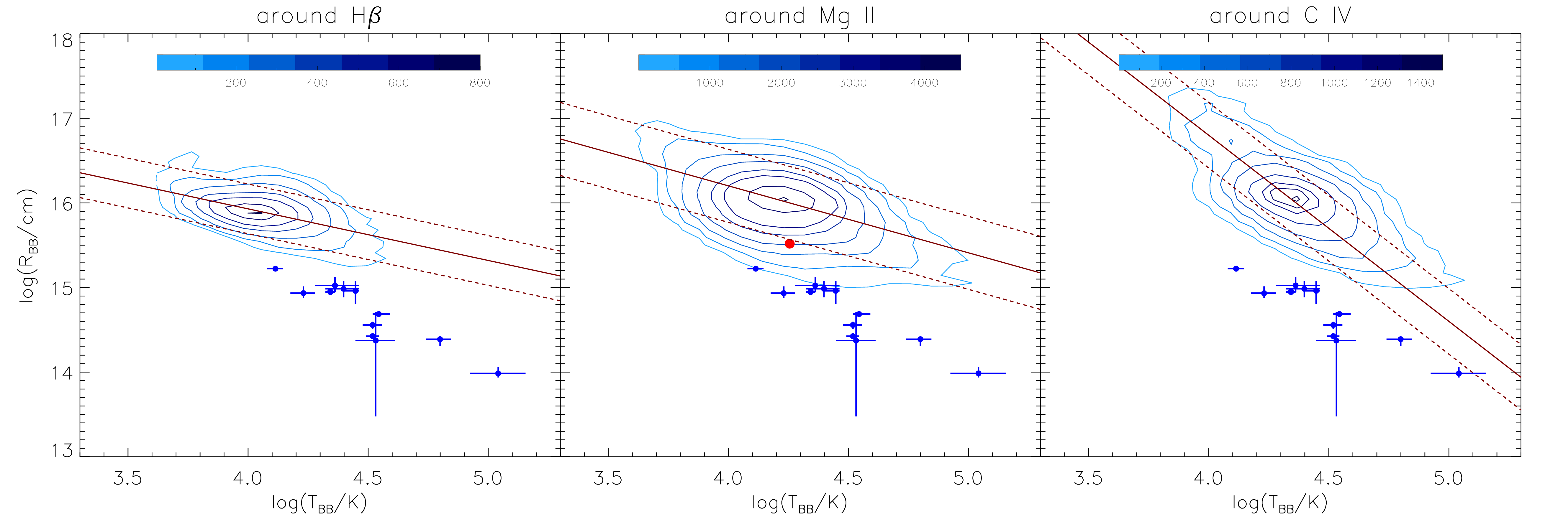}
\caption{Top panels show the best fitting results (solid red line) by the blackbody 
photosphere model to the power law described AGN continuum emissions (solid dark green line). 
In each top panel, the corresponding information of PLATE-MJD-FIBERID and redshift for 
the collected SDSS quasar is marked in title, the reported spectral index and the determined 
$T_{BB}$ and $R_{BB}$ are marked in top right corner. Bottom panels show the dependence of 
$R_{BB}$ on $T_{BB}$ of the SDSS quasars with reliable spectral indices. In each bottom 
panel, solid and dashed dark red lines show the best fitting results and the corresponding 
2.6RMS (99\%) scatters to the dependence for the quasars, solid circles plus error bars in 
blue show the results for the reported TDE candidates in \citet{mg19}, solid circle in red 
shows the results for the \obj. In each bottom panel, as the shown colorbar on the top, the 
contour levels in different colors show the numbers of quasars in the evenly divided 
40$\times$40 regions in the space of $\log(R_{BB})$ versus $\log(T_{BB})$.}
\label{main}
\end{figure*}

\begin{figure*}
\centering\includegraphics[width = 18cm,height=8cm]{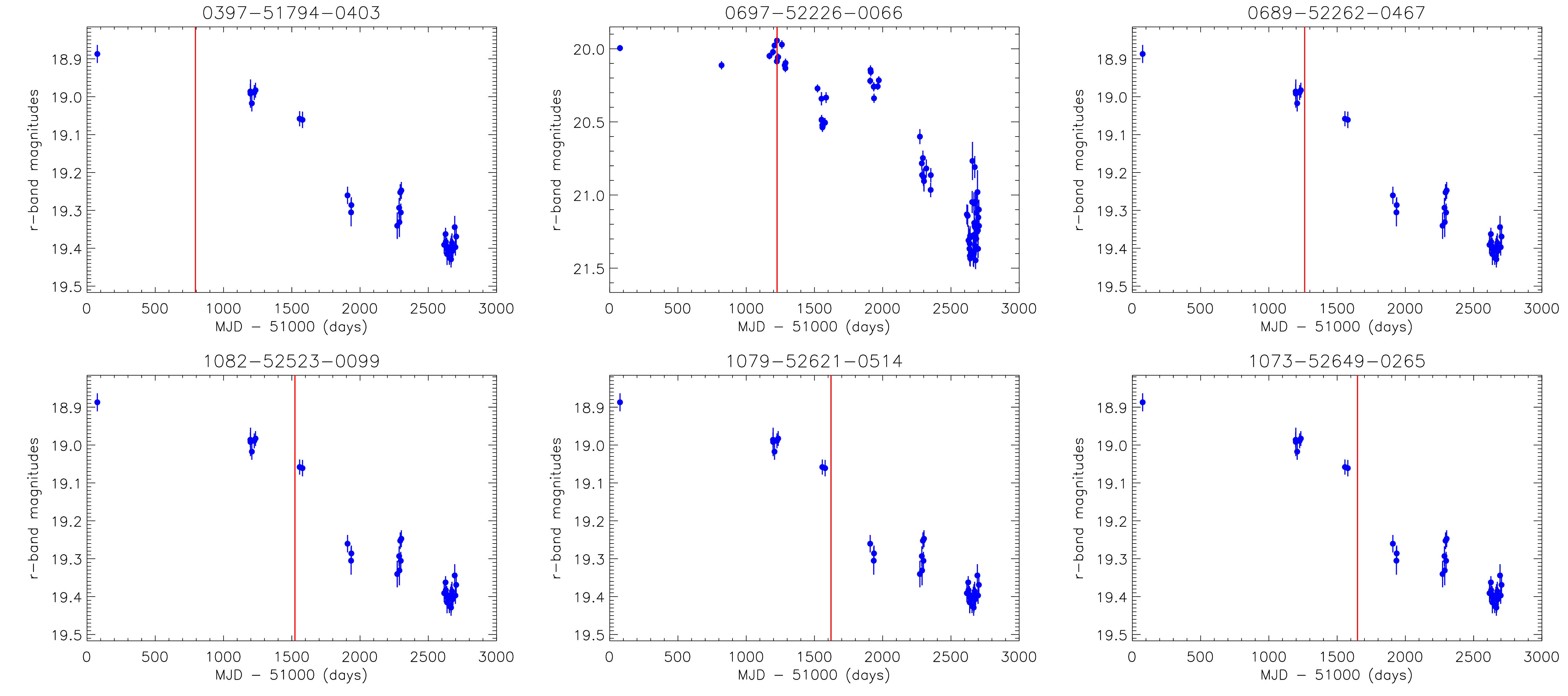}
\caption{The SDSS $r$-band light curves in observer frame of the six objects which could be 
potential candidates for TDEs but with no apparent information for peak brightness of the 
light curves. Vertical red line marks the MJD position for the SDSS spectrum with 
corresponding PLATE-MJD-FIBERID listed in title of each panel.}
\label{lmcs}
\end{figure*}

\begin{figure*}
\centering\includegraphics[width = 18cm,height=12cm]{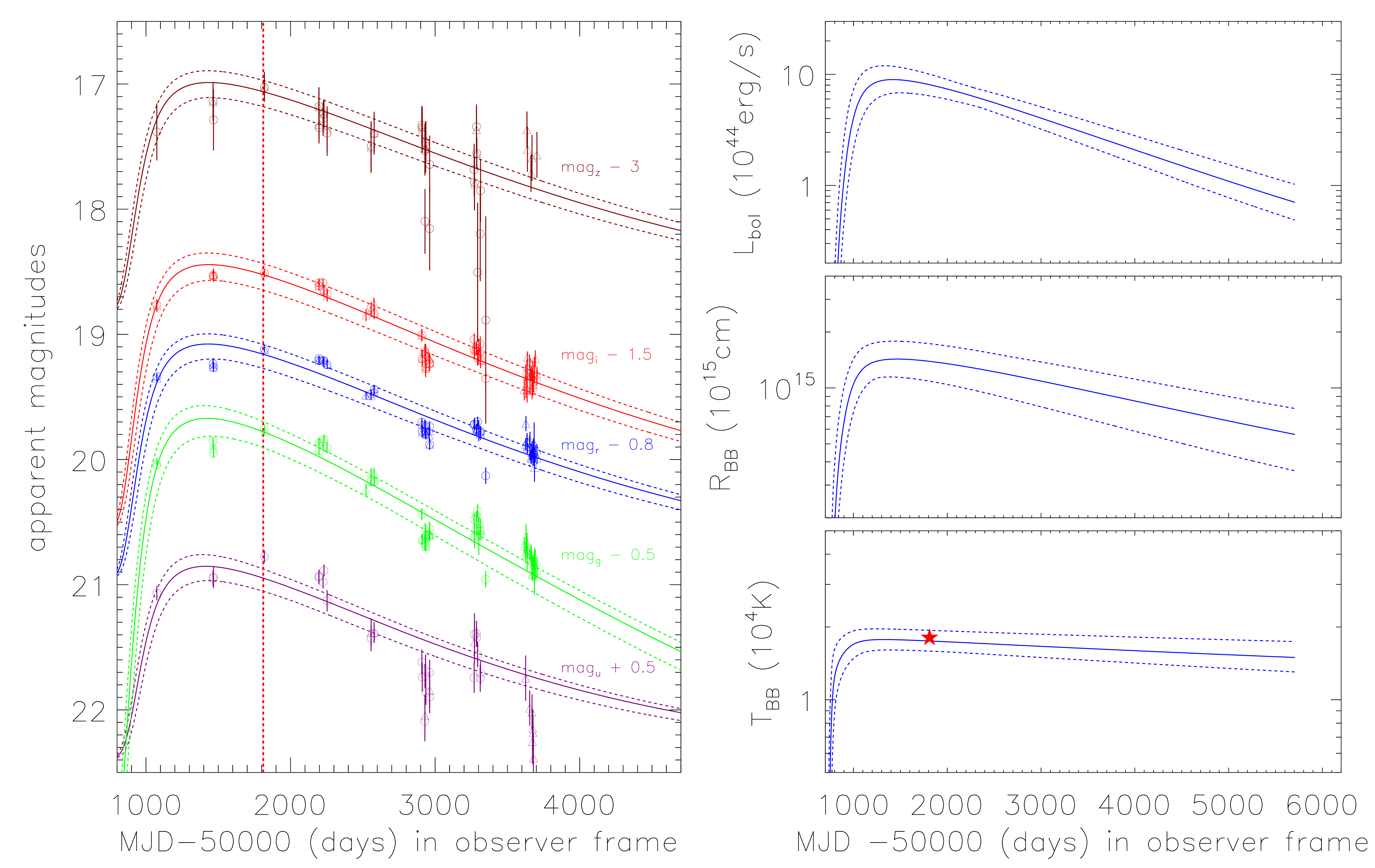}
\caption{Left panel shows the SDSS $ugriz$-band light curves and the theoretical TDE model 
determined best descriptions. In left panel, open circles and open triangles plus error bars 
in purple, in green, in blue, in red and in dark red show the data points from the PHOTOOBJALL 
database and from the Stripe82 database in the $u$-band, in the $g$-band, in the $r$-band, 
in the $i$-band, in the $z$-band, respectively. In left panel, solid lines in purple, in 
green, in blue, in red and in dark red show the TDE model determined best descriptions to 
the $ugriz$-band light curves, dashed lines in the same colors show the corresponding 
confidence bands to the best descriptions after accepted the $1\sigma$ uncertainties of the 
model parameters. In left panel, vertical dashed line in red marks the position of MJD=51811 
(the date for the SDSS spectrum of \obj). Right panels show the properties of the TDE model 
expected time dependent bolometric luminosity $L_{bol}$ (top right panel), photosphere 
radius $R_{BB}$ (middle right panel) and effective blackbody temperature $T_{BB}$ (bottom 
right panel) of \obj. In each right panel, dashed lines in blue show the corresponding 
confidence bands after accepted the $1\sigma$ uncertainties of the TDE model parameters of 
\obj. In the middle right panel and bottom right panel, the solid five-pointed star shows 
the corresponding results determined through the best descriptions to the continuum 
emissions of the SDSS spectrum of \obj~ by the simple blackbody photosphere model.}
\label{tde}
\end{figure*}

	Commonly, the long-term time-dependent TDEs expected unique variability pattern is 
the fundamental characteristic to detect TDE candidates. In this manuscript, starting from 
spectroscopic properties not from photometric variability properties, it is a try to check 
whether indicators can be estimated for detecting potential TDE candidates in quasars which 
have similar spectroscopic properties as TDE candidates. Therefore, the SDSS quasars with 
reported spectroscopic properties in \citet{sh11} are mainly considered. The manuscript is 
organized as follows. Section 2 presents the main results from SDSS spectra and the 
comparisons of $R_{BB}$ and $T_{BB}$ between SDSS quasars, leading to seven objects as TDE 
candidates. Section 3 mainly show the detailed discussions on the TDE candidate, the SDSS 
J230810-000021 (=\obj) at $z=1.16$, among the seven collected TDE candidates and provides 
discussions on properties of spectroscopic results and long-term photometric results to 
support the central TDE in \obj. Section 4 shows our main summary and conclusions. And in 
this manuscript, we have adopted the cosmological parameters of 
$H_{0}=70{\rm km\cdot s}^{-1}{\rm Mpc}^{-1}$, $\Omega_{\Lambda}=0.7$ and $\Omega_{\rm m}=0.3$.

\section{Blackbody photosphere model for continuum emissions of SDSS quasars}

	For the SDSS quasars in \citet{sh11}, the spectral indices, $\alpha_{H\beta}$, 
$\alpha_{MgII}$ and $\alpha_{CIV}$ have been measured and reported for continuum emissions 
around H$\beta$ within rest wavelength from 4400\AA~ to 5600\AA, for continuum emissions 
around Mg~{\sc ii} within rest wavelength from 2200\AA~ to 3100\AA, and for continuum 
emissions around C~{\sc iv} within rest wavelength from 1450\AA~ to 1700\AA. Accepted the 
reported spectra indices $\alpha$ and the corresponding continuum emission intensities 
$F_{\lambda_0}$ around 5100\AA, 3000\AA~ and 1350\AA~ in \citet{sh11}\footnote{The continuum 
luminosity is listed for each quasar in \citet{sh11}, however, combining with both redshift 
(to calculate the distance $D$ and continuum emission wavelength $\lambda_0$, it is easy to 
transform continuum luminosity $L_{\lambda_0}$ in units of erg/s to continuum emission 
intensity $F_{\lambda_0}$ in units of erg/s/${\rm cm^2}$/\AA~ by 
$F_{\lambda_0}=\frac{L_{\lambda_0}}{4\pi D^2\lambda_0}$.}, the pure continuum emissions 
without contaminations of emission lines can be simply obtained for each SDSS quasar, 
\begin{equation}
	F_{\lambda,qsos}~=~F_{\lambda_0}\lambda^\alpha
\end{equation}
as the shown three examples in top panels of Fig.~\ref{main} for three collected SDSS quasars. 
Here, not the pure power law described continuum emissions are shown in the top panels of 
Fig.~\ref{main}, but noises have been added with signal-to-noise to be randomly from 10 to 
50\footnote{Different values of signal-to-noise have few effects on our results shown in 
Fig.~\ref{main}.}, to consider additional uncertainties from subtractions of emission lines.

	Then, the blackbody photosphere model as discussed in \citet{mg19} is applied to 
describe each power law described continuum emission around H$\beta$ (Mg~{\sc ii} or C~{\sc iv}),
\begin{equation}
F_\lambda=\frac{2\pi Gc^2}{\lambda^5}\frac{1}{exp(hc/(k\lambda T_{BB}))-1}(\frac{R_{BB}}{D})^2
\end{equation}
with $G$, $c$, $h$, $k$ as the gravitation constant, the speed of light, the Planck constant, 
the Boltzmann constant, and $D$ as the distance (calculated by redshift) between the quasar 
and the observer, and $T_{BB}$ $R_{BB}$ as the photosphere temperature and the photosphere 
radius applied in the blackbody photosphere model. Top panels of Fig.~\ref{main} show the best 
descriptions to the power law described continuum emissions in the three SDSS quasars by the 
blackbody photosphere model. The applications of blackbody photosphere model can lead to 
estimated $R_{BB}$ and $T_{BB}$ for each continuum emission in each quasar. Then, bottom 
panels of Fig.~\ref{main} show the dependence of $R_{BB}$ on $T_{BB}$ based on the continuum 
emissions around H$\beta$, Mg~{\sc ii} and C~{\sc iv} for all the quasars (more than 120000) 
in \citet{sh11}, respectively.

	Based on the measured $R_{BB}$ and $T_{BB}$ for the SDSS quasars, through the Least 
Trimmed Squares (LTS) robust technique \citep{cap13, mm17}, the linear fitting results and 
the corresponding 2.6RMS (99\%) scatters shown in the bottom panels can be described by 
\begin{equation}
\begin{split}
	\log(\frac{R_{BB}}{\rm cm})&=(18.37\pm0.38)-(0.61\pm0.09)\log(\frac{T_{BB}}{\rm K}) \\
	\log(\frac{R_{BB}}{\rm cm})&=(19.37\pm0.21)-(0.79\pm0.05)\log(\frac{T_{BB}}{\rm K}) \\
	\log(\frac{R_{BB}}{\rm cm})&=(25.60\pm0.48)-(2.20\pm0.11)\log(\frac{T_{BB}}{\rm K}) \\
\end{split}
\end{equation}
with RMS scatters about 0.113, 0.166 and 0.149 for the 17485 SDSS quasars with reliable 
$\alpha_{H\beta}$, the 79343 SDSS quasars with reliable $\alpha_{MgII}$, and the 26325 SDSS 
quasars with reliable $\alpha_{CIV}$, respectively.

	Besides the results from SDSS quasars, the reported $R_{BB}$ and $T_{BB}$ of the 
TDE candidates listed in \citet{mg19} are also shown in the bottom panels of Fig.~\ref{main}, 
with the $R_{BB}$ and $T_{BB}$ as the values related to the peak brightness of the TDEs 
expected light curves. It is clear that there are distinct two regions for the SDSS quasars 
and the TDEs lying in the space of $R_{BB}$ versus $T_{BB}$. In other words, the large gaps 
between the SDSS quasars and the TDEs in the space of $R_{BB}$ versus $T_{BB}$ could provide 
further clues to distinguish quasars from TDEs.

	Through the shown results in Fig.~\ref{main}, it is a good try to test whether the 
SDSS quasars (as outliers) locating below the lower boundary of the 2.6RMS scatters should 
be potential candidates for TDEs. There are 197, 1801 and 111 SDSS quasars locating below 
the lower boundary of the 2.6RMS scatters in the bottom left panel, bottom middle panel, and 
bottom right panel, respectively. Meanwhile, in order to check their long-term variability 
properties, the collected outliers are matched with the SDSS Stripe82 database, and there 
are 31, 358 and 13 outliers covered by the SDSS Stripe82 region, respectively. Then, through 
the reported light curves from the SDSS Stripe82 database 
(\url{http://das.sdss.org/value_added/stripe_82_variability/SDSS_82_public/}) with detailed 
descriptions in \citet{bv08}, the long-term variability properties of the 402 outliers have 
been carefully checked.

	As described in the Introduction, the shown results for the TDEs in the bottom 
panels of Fig.~\ref{main} are the values around the epoch for the TDEs expected peak 
brightness. Therefore, among the 402 outliers, we try to find candidates of which SDSS 
spectra have been observed around the epoch for the peak brightness of their light curves 
having unique variability patterns with a steep rise followed by a smooth declined trend 
expected by theoretical TDEs model. Then, there is one object, \obj~ at $z=1.16$, with its 
SDSS spectrum (PLATE-MJD-FIBERID = 0381-51811-0461) observed around the epoch for the peak 
brightness of its light curve, providing the better chance to compare properties of $R_{BB}$ 
and $T_{BB}$ from spectroscopic results and from theoretical TDE model expected results in 
the following section. And properties of both spectroscopic and photometric results could 
provide a new method to support the central TDE in the \obj.

	Before ending the section, one point should be noted. Among the 402 outliers, there 
are six another objects with their light curves having smooth decline trends, indicating clues 
for central TDEs. However, due to loss of data points around peak brightness of the light 
curves, it is hard to determine the properties of $R_{BB}$ and $T_{BB}$ through applications 
of theoretical TDE model to describe the long-term light curves, and then there is no way to 
compare the properties of $R_{BB}$ and $T_{BB}$ from spectroscopic results and from theoretical 
TDE model expected results. Therefore, in the following sections, properties of photometric 
light curves and SDSS spectrum are mainly considered in the \obj. And there are no further 
discussions of the six objects any more in this manuscript, but showing the light curves of 
the six objects in Fig.~\ref{lmcs}. Detailed discussions on the six objects should be given 
in one following being prepared manuscript in the near future, to provide an independent 
method to determine the loss time information of peak brightness in the light curves related 
to central TDEs.

\section{Photometric and spectroscopic results of \obj}

	Through the SDSS Stripe82 database as described in \citet{bv08} combined with the 
SDSS public database of PHOTOOBJALL\footnote{detailed descriptions on the database can be found 
in \url{https://cas.sdss.org/dr16/en/help/browser/browser.aspx?cmd=description+PhotoObjAll+U}} 
(with THINGID=94258814), the 7.2-year-long SDSS $ugriz$-band light curves of the \obj~ can be 
collected and shown in left panel of Fig.~\ref{tde}, with MJD from 51075 (Sep. 1998) to 53705 
(December 2005). Here, the data points from the PHOTOOBJALL being consistent with the data 
points from the Stripe82 database, especially in the $gri$-band light curves, can be applied 
to support that the shown light curves are reliable enough in the \obj.

	Due to the unique variability pattern with a rise followed by a smooth decline trend 
in the long-term variability of \obj, theoretical TDE model is firstly considered whether the 
photometric variability can be described. More recent detailed descriptions (and corresponding 
public codes of TDEFIT/MOSFIT) on the theoretical TDE model accepted in this manuscript can 
be found in \citet{gr13, gm14, mg19}. The theoretical TDE model can be applied as follows to 
describe the light curves of \obj~ based on our created templates of viscous-delayed accretion 
rates, similar as what we have recently done in \citet{zh22, zh22b, zh22c, zh23a, zh23b, zh24a, 
zh24b}.

	For standard TDEs cases with a solar-like main-sequence star (stellar mass 
$M_*=1M_\odot$) disrupted by a supermassive black hole (SMBH) (mass $M_{BH}=10^6M_\odot$) with 
different impact parameters $\beta$, there are time dependent fallback material rates 
$\dot{M}_{f}$  provided by the public TDEFIT/MOSFIT code. Then, as described in \citet{mg19}, 
templates of the viscous-delayed accretion rates $\dot{M}_{a}$ can be created  
with different viscous-delay time scales ($T_{vis}$ in units of years)
\begin{equation}
\dot{M}_{a}(T_{vis})~=~\frac{exp(-t/T_{vis})}{T_{vis}}\int_{0}^{t}exp(t'/T_{vis})\dot{M}_{f}dt'
\end{equation}. 
The calculated $\dot{M}_{a}$ depends on the same time information $t_a$ for the 
$\dot{M}_{f}$. Then, for common TDE cases with $M_{\rm BH}$ and $M_{*}$ different from 
$10^6{\rm M_\odot}$ and $1{\rm M_\odot}$, as discussed in \citet{gr13, mg19}, corresponding 
actual viscous-delayed accretion rates $\dot{M}$ and the corresponding time information can 
be created by 
\begin{equation}
\begin{split}
&\dot{M} = M_{\rm BH,6}^{-0.5}\times M_{\star}^2\times R_{\star}^{-1.5}\times\dot{M}_{a}(T_{vis}, \beta) \\
&t = (1+z)\times M_{\rm BH}^{0.5}\times M_{\star}^{-1}\times R_{\star}^{1.5} \times t_{a}(T_{vis}, \beta)
\end{split}
\end{equation},
where $M_{\rm BH,6}$, $M_{\star}$, $R_{\star}$ and $z$  are central BH mass in units of 
${\rm 10^6M_\odot}$, stellar mass in units of ${\rm M_\odot}$, stellar radius in units of 
${\rm R_{\odot}}$ and redshift, respectively. Here, the mass-radius relation discussed in 
\citet{tp96} has been accepted in the manuscript for main-sequence stars, in order to ignore 
effects of coupled parameters of $M_{\star}$ and $R_{\star}$.

	Based on the TDE model expected time-dependent actual accretion rates $\dot{M}$, the 
time-dependent output emission spectrum in the rest frame can be determined by the simple 
blackbody photosphere model as discussed in \citet{gm14, mg19}, 
\begin{equation}
\begin{split}
	&F_\lambda(t)=\frac{2\pi Gc^2}{\lambda^5}\frac{1}{exp(hc/(k\lambda T_{BB}(t)))-1}(\frac{R_{BB}(t)}{D})^2 \\
	&R_{BB}(t) = R_0\times a_p(\frac{\epsilon\dot{M}c^2}{1.3\times10^{38}M_{\rm BH}/{\rm M_\odot}})^{l_p} \\ 
	&T_{BB}(t)=(\frac{\epsilon\dot{M}c^2}{4\pi\sigma_{SB}R_{BB}^2})^{1/4} \ \ \ \ \ \ 
	a_p = (G M_{\rm BH}\times (\frac{t_p}{\pi})^2)^{1/3}	
\end{split}
\end{equation}
with $D$ as the distance to the earth calculated by the redshift $z$, $k$ as the Boltzmann 
constant, $T_{BB}(t)$ and $R_{BB}(t)$ as the time-dependent effective temperature and radius 
of the photosphere, respectively, and $\epsilon$ as the energy transfer efficiency smaller 
than 0.4, $\sigma_{SB}$ as the Stefan-Boltzmann constant, and $t_p$ as time information of the 
peak accretion.

	Based on the TDE model expected emission spectrum $F_\lambda(t)$ convoluted with the 
transmission curves of SDSS $ugriz$ filters, the time-dependent SDSS $ugriz$-band magnitudes 
$mag_{u,~g,~r,~i,~z}(t, TDE)$ of \obj~ can be described by the standard TDE model. Moreover, 
there are five additional parameters $mag_{0}(u,~g,~r,~i,~z)$ (not time dependent parameters), 
applied to describe the contributions of the host galaxy (or the none-variability component) 
to observed variability in SDSS $ugriz$ bands. Then, the observed time dependent magnitudes 
$mag_{u,~g,~r,~i,~z}(t, obs)$ in each band can be described by
\begin{equation}
	10^{\frac{mag(t, obs)}{-2.5}} = 10^{\frac{mag(t, TDE)}{-2.5}} + 10^{\frac{mag_{0}}{-2.5}}
\end{equation}
. Through the Levenberg-Marquardt least-squares minimization technique (the known MPFIT 
package) \citep{mc09}, the best descriptions can be determined to the SDSS $ugriz$-band 
light curves of \obj. The corresponding determined TDE model parameters (with polytropic 
index $\gamma=4/3$ for the disrupted star) and the corresponding $1\sigma$ uncertainties 
from the covariance matrix are: $\log(M_{\rm BH,6})\sim1.73\pm0.11$, 
$\log(M_\star/{\rm M_\odot})\sim0.78\pm0.05$ (the corresponding stellar radius 
$\log(R_\star/{\rm R_\odot})\sim0.44\pm0.03$), $\log(\beta)\sim0.21\pm0.02$, 
$\log(T_{vis})\sim-0.69\pm0.12$, $\log(\epsilon)\sim-1.59\pm0.05$, 
$\log(R_0)\sim-0.40\pm0.05$, $\log(l_p)\sim-0.39\pm0.03$, $\log(mag_0(u))\sim1.34\pm0.01$, 
$\log(mag_0(g))\sim1.37\pm0.05$, $\log(mag_0(r))\sim1.34\pm0.01$, 
$\log(mag_0(i))\sim1.34\pm0.01$, $\log(mag_0(z))\sim1.33\pm0.01$. Left panel of Fig.~\ref{tde} 
shows the TDE model determined best-fitting results and the corresponding confidence bands 
after accepted the 1$\sigma$ uncertainties of the model parameters.

	Before proceeding further, three points are noted. For the first point, the TDE 
model determined BH mass is around $5.4\times10^7{\rm M_\odot}$ 
($\log(M_{\rm BH,6})\sim1.73\pm0.11$,) in \obj. It is necessary to check whether so large BH 
mass is reasonable for the expected central TDE in \obj. Based on the definition of Hills 
limit $M_H$ for TDEs around Schwarzschild SMBHs (see equation 6 in \citealt{yr23}),
\begin{equation}
	M_{H} = 1.1\times10^8 (\frac{M_{\star}}{\rm M_\odot})^{-1/2}(\frac{{R_\star}}{\rm R_\odot})^{3/2} {\rm M_\odot}
\end{equation},
accepted the determined stellar parameters of $M_\star\sim6.03{\rm M_\odot}$ and 
$R_\star\sim2.77{\rm R_\odot}$ for the TDE candidate in SDSS J2308, the Hills limit about 
$\log(M_H/{\rm 10^6M_\odot})\sim2.71$ is about ten times larger than the TDE model determined 
BH mass $\log(M_{BH}/{\rm 10^6M_\odot})\sim1.73$ in SDSS J2308. Therefore, the TDE model 
determined BH mass is reasonable.

\begin{figure}
\centering\includegraphics[width = 8cm,height=5cm]{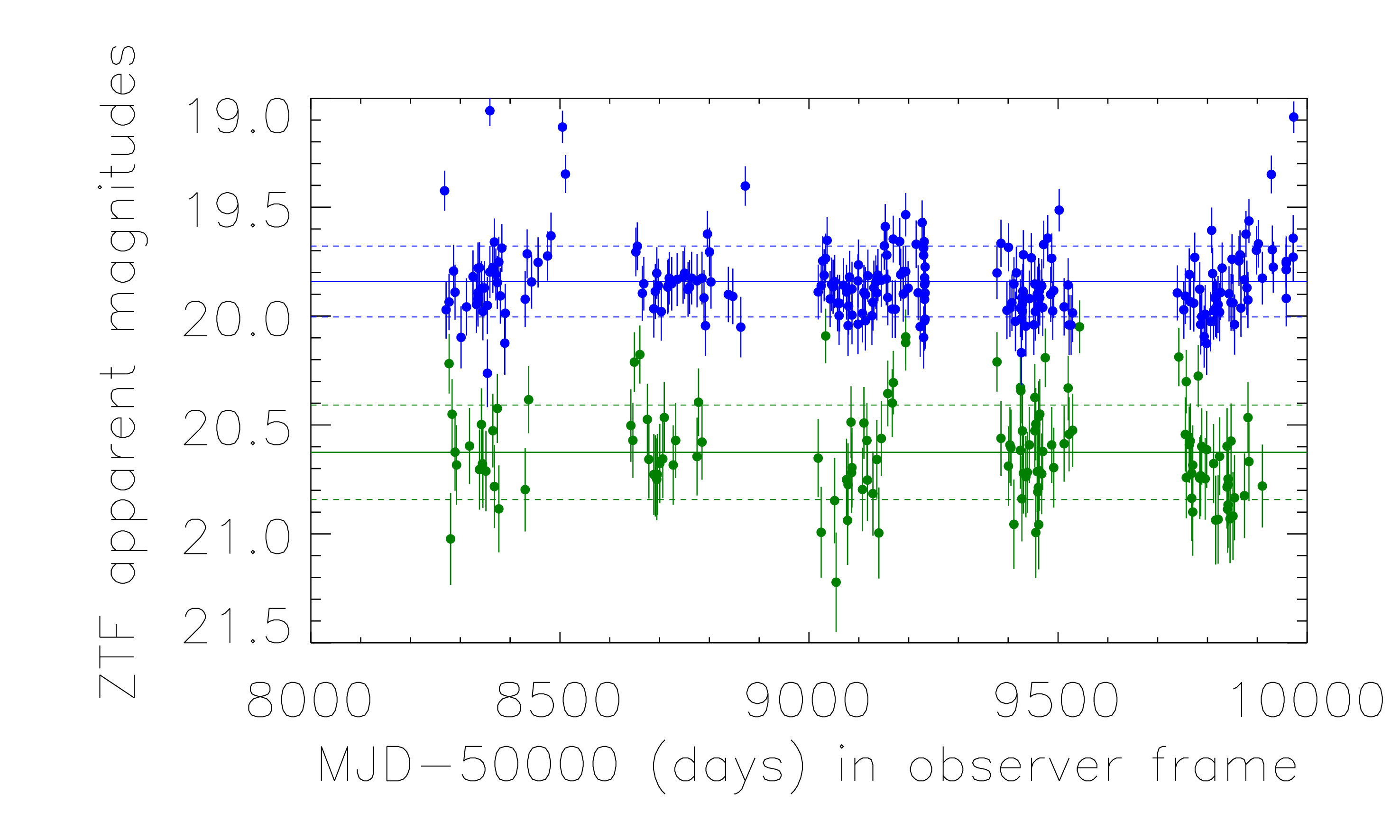}
\caption{The ZTF g/r-band light curves in observer frame. Solid circles plus error bars in 
dark green and in blue show the data points in the g-band light curve and the r-band light 
curve, respectively. Solid and dashed lines in dark green and in blue mark the mean value 
and the corresponding 1RMS scatters of the g-band light curve and the r-band light curve, 
respectively.}
\label{ztf}
\end{figure}

	For the second point, besides the long-term light curves from the SDSS Stripe82 
database, we have also checked the variability properties of SDSS J2308 through the public 
sky survey projects of CSS (Catalina Sky Survey) \citep{dr09}, ASAS-SN (All-Sky Automated 
Survey for Supernovae) \citep{sp14, ks17} and ZTF (Zwicky Transient Facility) \citep{bk19, 
ds20}. Available long-term g/r-band light curves of SDSS J2308 can be collected from the 
ZTF project. And there are no available light curves detected in the CSS or the ASAS-SN 
projects. The ZTF g/r-band light curves are shown in Fig.~\ref{ztf} with MJD-50000 in 
observer frame from 8268 (28th, Jun., 2018) to 10314 (4th, Jan., 2024). It is apparent that 
there are none apparent variability in the ZTF light curves with time durations longer than 
5.6 years in observer frame. Therefore, the optical flare in SDSS J2308 in the light curves 
provided by SDSS Stripe82 database can be confirmed to be unique enough, and to be very 
different from the intrinsic AGN variability as common features in different epochs.

	For the third point, as the shown results in Fig.~\ref{tde}, the time duration of 
the flare related to an assumed TDE is about 10 years, which are longer than the common 
time durations (about tens to a few hundreds of days) of the reported TDE candidates as 
discussed in \citet{mg19, vh20, vg21, yr23}, etc.. However, the longer time duration about 
10 years of the optical flare related to an assumed TDE in SDSS J2308 can be simply expected 
as follows. Considering different parameters of redshift and stellar parameters of disrupted 
stars and BH masses, based on the scaling relation shown in equation 5, the ratio $T_{R}$ 
of time duration between SDSS J2308 and the other reported optical TDE candidates can be 
described by
\begin{equation}
T_R~\sim~\frac{1+z}{1+z_0}(\frac{M_{BH}}{M_{BH0}})^{0.5}(\frac{M_{*}}{M_{*0}})^{-1}
	(\frac{R_{*}}{R_{*0}})^{1.5}\frac{t_a(\beta, T_{vis})}{t_{a}(\beta_0, T_{vis0})}
\end{equation}
with $z=1.16$, $M_{BH}\sim5.4\times10^7{\rm M_\odot}$, $M_{*}\sim6.03{\rm M_\odot}$, 
$R_{*}\sim2.77{\rm R_\odot}$, $t_a(\beta=1.6, T_{vis}=72days)$ as the parameters applied 
in SDSS J2308 and with $M_{BH0}$, $M_{*0}$, $R_{*0}$, $t_{a}(\beta_0)$ as the parameters 
applied in the common TDE candidates. Among the reported TDE candidates with determined TDE 
model parameters determined in \citet{mg19}, the D3-13 at $z_0\sim0.3698$ \citep{gb08} has 
the model parameters of $M_{BH0}\sim3\times10^7{\rm M_\odot}$, $M_{*0}\sim7{\rm M_\odot}$, 
$R_{*0}\sim3.02{\rm R_\odot}$ and $t_a(\beta_0\sim1.8, T_{vis0}\sim36days)$ which are similar 
as the parameters of SDSS J2308. Then, compared with time duration about 1200days (see results 
in Figure 1 in \citealt{mg19}), the expected time duration of SDSS J2308 should be $T_R\sim3$ 
times of the time duration in D3-13, i.e., the expected time duration to be about 3600days 
(3$\times$1200days) in SDSS J2308 which are well consistent with the shown time duration 
about 10years for the optical flares in SDSS J2308. Even compared with the time durations 
of the other TDE candidates, considering the different TDE model parameters, the longer 
time duration in the optical flare in SDSS J2308 can also be expected.

	Moreover, based on the TDE model determined parameters, the time dependent bolometric 
luminosity $L_{bol}$ in \obj~ can be estimated by the determined actual viscous-delayed 
accretion rates $\dot{M}$ in equation (5), the time dependent effective blackbody temperature 
$T_{BB}(t)$ and photosphere radius $R_{BB}(t)$ in \obj~ can be estimated by equation (6), 
shown in right panels of Fig.~\ref{tde}. Comparing with the results for the reported TDEs 
in \citet{mg19} (see their Fig.~8), the time dependent $L_{bol}$, $R_{BB}$ and $T_{BB}$ in 
\obj~ are common, indicating model parameters of the assumed central TDE are reasonable in \obj.

\begin{figure}
\centering\includegraphics[width = 8cm,height=10cm]{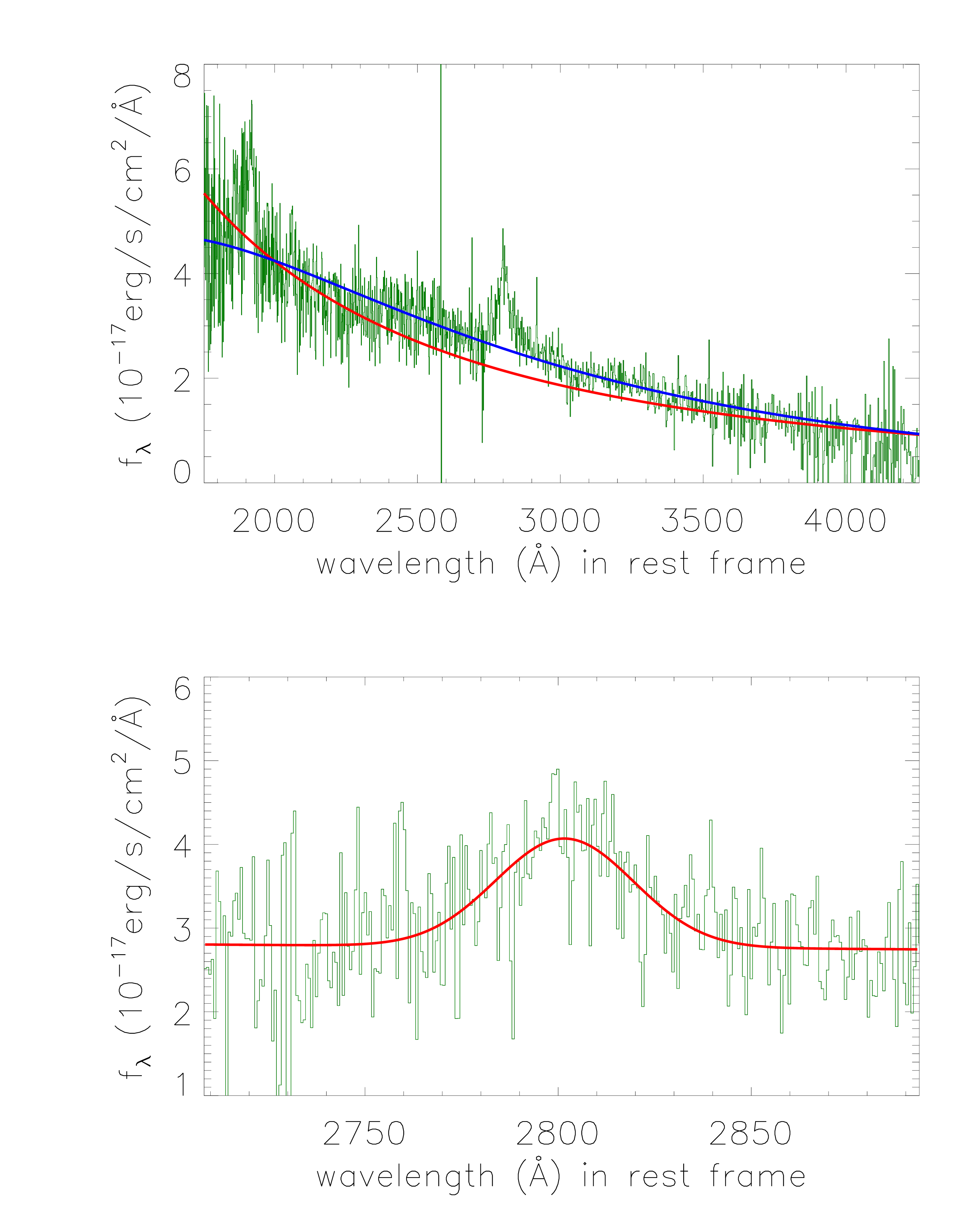}
\caption{Top panel shows the SDSS spectrum of \obj~ with PLATE-MJD-FIBERID=0381-51811-0461. 
Solid red line shows the determined power law continuum emissions, solid blue line shows the 
blackbody photosphere model determined SED. Bottom panel shows the broad Mg~{\sc ii} line, 
and the best fitting results (solid red line) by one Gaussian function plus a power law component.}
\label{spec}
\end{figure}

	Besides the long-term photometric light curves, the SDSS spectrum 
(PLATE-MJD-FIBERID=0381-51811-0461) of \obj~ is collected and shown in top panel of 
Fig.~\ref{spec} with apparent broad Mg~{\sc ii}$\lambda2800$\AA~ and broad 
C~{\sc iii}]$\lambda1905$\AA~ emission lines leading to the determined redshift about 1.16. 
Continuum emissions in \obj~ can be describe by a power law function 
$f_\lambda~\propto~\lambda^{-2.02\pm0.05}$ in the rest frame (shown as solid blue line), 
leading the continuum luminosity at 3100\AA~ in the rest frame to be 
$\lambda L_{3100}\sim3.73\times10^{44}{\rm erg/s}$. The broad Mg{\sc ii} emission line is 
also shown in the bottom panel of Fig.~\ref{spec}, which can be described by a broad Gaussian 
function with the second moment $\sigma\sim17.9\pm1.8$\AA~ and the line flux about 
$(58.1\pm6.5)\times10^{-17}{\rm erg/s/cm^2}$ in the rest frame.

	As a high redshift broad emission line quasar classified by the SDSS pipeline, there 
are few contributions of host galaxies to the SDSS spectrum of \obj~ at MJD=51811 around the 
time for the peak brightness of the light curves shown in left panel of Fig.~\ref{tde}. 
Therefore, it is a better chance to check whether a pure blackbody photosphere model can be 
applied to describe the continuum emission properties in the SDSS spectrum of \obj~ shown in 
top panel of Fig.~\ref{spec}. Through the simple blackbody photosphere model, the SED (spectral 
energy distributions) with rest wavelength smaller than 2600\AA~ or larger than 3200\AA~ (to 
ignore effects of broad Mg~{\sc ii} emission line) of \obj~ can be described through the 
Levenberg-Marquardt least-squares minimization technique, shown as solid blue line in top 
panel of Fig.~\ref{spec} with $\chi^2/dof\sim1.36$, with determined blackbody temperature 
about 18070$\pm$140K and photosphere radius about $(3.3\pm0.1)\times10^{15}{\rm cm}$ which 
have been marked in the bottom middle panel of Fig.~\ref{main}, consistent with the TDE model 
determined results shown in right panels of Fig.~\ref{tde}, providing further clues to 
support the central TDE in \obj.

\begin{figure}
\centering\includegraphics[width = 8cm,height=15cm]{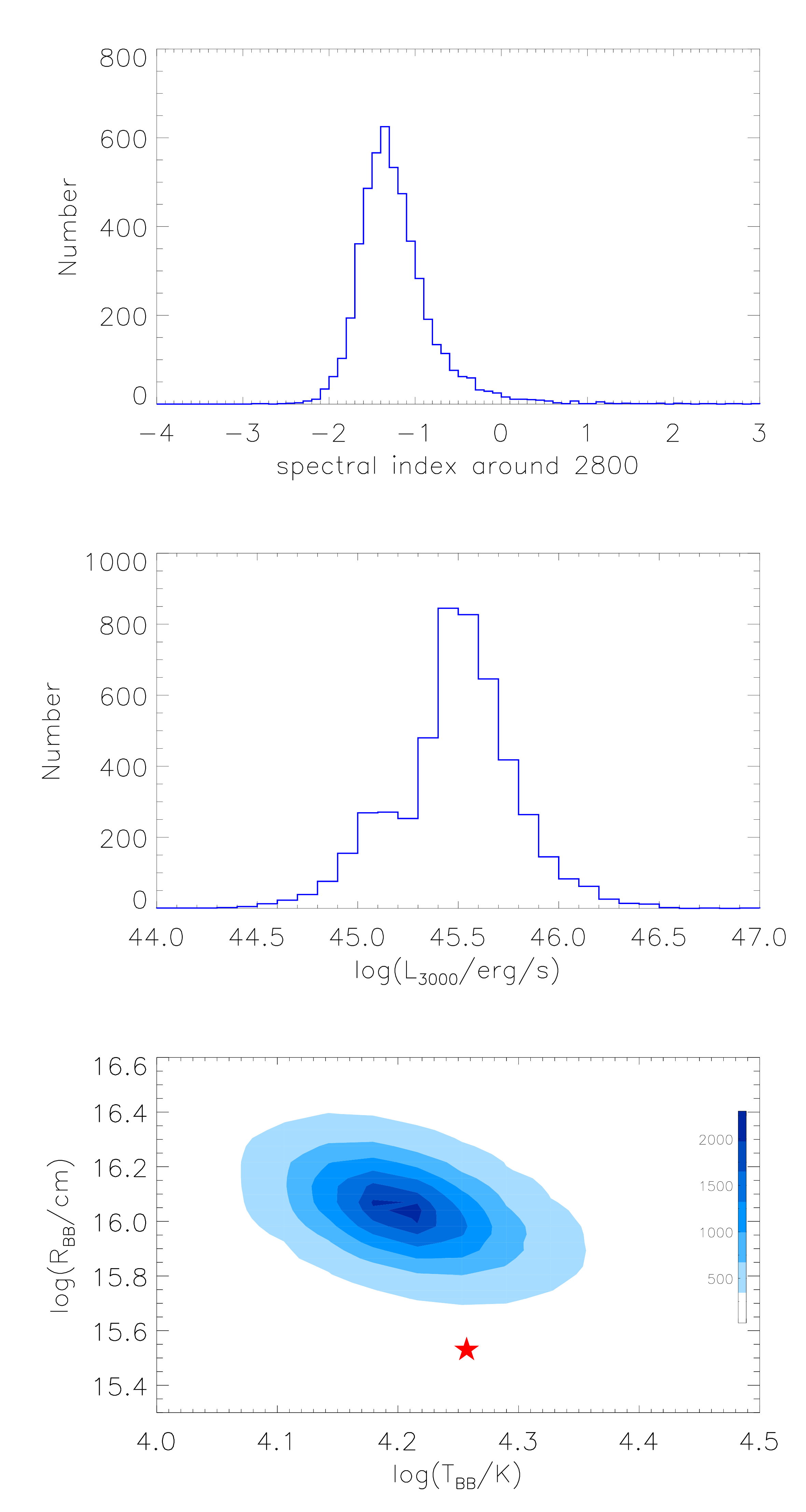}
\caption{Top panel and middle panel show distributions of the spectral index $\alpha$ around 
broad Mg~{\sc ii} emission line and the continuum luminosity at 3000\AA~ of the collected 
4937 quasars. Bottom panel shows the two-dimensional distributions of $T_{BB}$ and $R_{BB}$ 
of the 100000 artificial continuum emissions created by the distributions of $\alpha$ and 
$L_{3000}$. In bottom panel, solid five-pointed star marks the position for \obj. In bottom 
panel, as the shown colorbar on the right, the contour levels in different colors show the 
numbers of cases in the evenly divided 40$\times$40 regions in the space of $\log(R_{BB})$ 
versus $\log(T_{BB})$.}
\label{mcbb}
\end{figure}

	Furthermore, a simple method is applied to determine a probability of such a blackbody 
photosphere model described continuum emissions in \obj~ related to normal quasars. In the 
database of \citet{sh11}, there are 4937 quasars with redshift larger than 1.1 and smaller 
than 1.2 (redshift 1.16 of \obj). Distributions of the spectral index $\alpha$ around 2800\AA~ 
and the continuum luminosity $L_{3000}$ at 3000\AA~ in rest frame of the 4937 quasars are 
collected and shown in top panel and middle panel of Fig.~\ref{mcbb}. Then, based on the 
distributions of $\alpha$ and $L_{3000}$, 100000 artificial continuum emissions can be created by 
\begin{equation}
f_\lambda = \frac{L_{3000}}{f_{sca}}~\times~(\frac{\lambda}{3000\textsc{\AA}})^{\alpha}
\end{equation}, 
with $f_{sca}\sim7.54\times10^{57}cm^2\textsc{\AA}$ as the value applied to transform 
continuum luminosity to emission intensity in units of $10^{-17}erg/s/cm^2/\textsc{\AA}$ at 
redshift 1.16 (the redshift of \obj), and $\lambda$ as the same wavelength as that of \obj. 
Then, for each artificial continuum emission, the blackbody photosphere model is applied to 
determine the $T_{BB}$ and $R_{BB}$. Bottom panel of Fig.~\ref{mcbb} shows the distributions 
of $T_{BB}$ and $R_{BB}$ of the 100000 artificial continuum emissions. Considering the 
determined $T_{BB}$ and $R_{BB}$ within ranges of $18070-1400\le T_{BB}/K \le 18070+1400$ (18070 
and 1400 as the determined value and 10 times of the 1$\sigma$ uncertainty of $T_{BB}/K$ of 
\obj) and $339-41\le R_{BB}/10^{15}cm\le 339+41$ (339 and 41 as the determined value and 10 
times of the 1$\sigma$ uncertainty of $R_{BB}/10^{15}cm$ of \obj), there are only 11 
artificial continuum emissions, indicating the probability about $1.1\times10^{-4}$ that the 
continuum emissions with determined $T_{BB}\sim18070K$ and $R_{BB}\sim3.39\times10^{15}cm$ 
in \obj~ are related to continuum emissions of normal quasars. In other words, the confidence 
level is higher than 4$\sigma$ that the continuum emissions are tightly related to the 
central TDE in \obj.

\begin{figure}
\centering\includegraphics[width = 8cm,height=15cm]{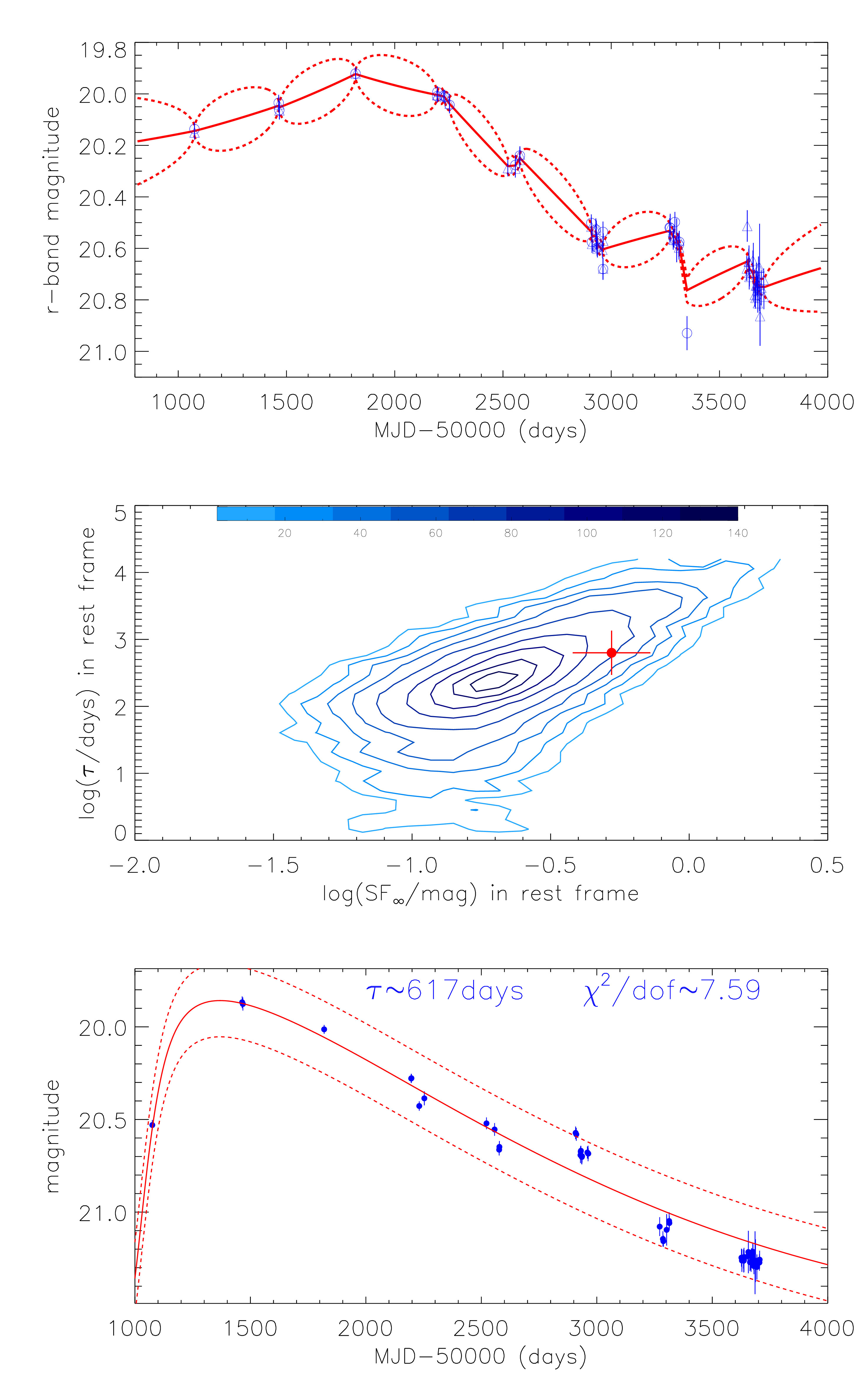}
\caption{Top panel shows the JAVELIN code determined best descriptions (solid red line) and 
the corresponding $1\sigma$ confidence bands (dashed red lines) to the $r$-band light curve 
(symbols in blue as those in left panel of Fig.~\ref{tde}) of \obj. Middle panel shows the 
dependence of $\log(\tau)$ on $\log(SF_\infty)$ in rest frame for the SDSS quasars (in contour) 
and the \obj~ (solid circle plus error bars in red). In middle panel, as the shown colorbar 
on the top, the contour levels in different colors show the numbers of quasars in the evenly 
divided 30$\times$30 regions in the space of $\log(\tau)$ versus $\log(SF_\infty)$. Bottom 
panel shows an example of CAR simulating light curve (solid circles plus error bars) 
described (solid red line) by theoretical TDE model and the corresponding confidence bands 
after accepted the $1\sigma$ uncertainties of the model parameters. In top corner of 
bottom panel, the applied value $\tau$ and the determined $\chi^2/dof$ to the TDE model 
determined best fitting results are marked.}
\label{car}
\end{figure}

	Before ending the section, three additional points should be noted. First and 
foremost, considering broad Mg~{\sc ii} emission line, the properties of virial BH mass 
in \obj~ are simply discussed as follows. Through measured line parameters of the broad 
Mg~{\sc ii} emission lines, central virial BH mass as discussed in \citet{pf04, rh11, 
sh11}can be estimated as $2.88\times10^8{\rm M_\odot}$ by
\begin{equation}
\log(\frac{M_{BH}}{{\rm M_\odot}})=0.86+0.5\log(\frac{\lambda L_{3100}}{{\rm 10^{44}erg/s}}) + 2\log(\frac{FWHM}{{\rm km/s}})
\end{equation}
with $FWHM\sim4540{\rm km/s}$ ($2.35\sigma$) (full width at half maximum) of the broad 
Mg~{\sc ii} line in rest frame. The estimated virial BH mass is simply consistent with the 
reported BH mass $5.4_{-2.6}^{+5.1}\times10^8{\rm M_\odot}$ in \citet{sh11}. The virial BH 
mass is about 5.4 times larger than the TDE model determined value. However, considering 
the large uncertainties in virial BH mass, it is hard to confirm the virial BH mass different 
from the TDE model determined BH mass. Further efforts are necessary to estimate/measure 
the central BH mass of \obj~ by another independent methods. However, if accepted the larger 
virial BH masses in \obj, the assumed central TDE could be preferred in \obj, after 
considering the non-virial dynamic broad emission line clouds related to TDEs debris 
nearer to the central SMBH in \obj, as shown in the TDE candidate in SDSS J0159 discussed 
in \citet{zh23b} and in the TDE candidate ASASSN-14li discussed in \citet{ht16} with 
non-virial dynamic properties of broad line emission clouds related to TDEs debris.

	Besides, simple discussions are given on whether the long-term variability of \obj~ 
are not related to a central TDE but related to intrinsic AGN activity in \obj. As shown 
in \citet{mi10}, accepted the Continuous AutoRegressive (CAR) process \citep{kbs09, mv19, 
sr22} or damped random walk (DRW) process \citep{koz10, zk13, zk16}, long-term variability 
of the SDSS quasars covered in the Stripe82 region have been carefully analyzed, and leading 
to the dependence of variability timescale $\tau$ on long term variability structure function 
$SF_{\infty}$. Then, based on the JAVELIN code \citep{zk13, zk16} applied to the SDSS 
$r$-band\footnote{Similar parameters of $\tau$ and $SF_{\infty}$ can be obtained through 
the other band light curves of \obj} light curve of \obj~ as shown in the top panel of 
Fig.~\ref{car}, the determined $\tau$ and $SF_{\infty}$ in rest frame can be determined 
to be $\log(\tau/days)\sim2.76_{-0.29}^{+0.36}$ and 
$\log(SF_{\infty}/mag)\sim-0.28_{-0.11}^{+0.16}$. Then, middle panel of Fig.~\ref{car} 
shows properties of \obj~ in the space of $\tau$ versus $SF_{\infty}$ (same as the Fig.~3 
in \citealt{mi10}). Comparing with the normal SDSS quasars as discussed in \citet{mi10}, 
the \obj~ does not have totally different locations from those of the quasars in the 
space of $\tau$ and $SF_{\infty}$. Therefore, through the CAR process as described 
in \citet{kbs09}, 
\begin{equation}
\dif LMC_t~=~\frac{-1}{\tau}LMC_t~\dif~ t~+~\sigma_*~\sqrt{\dif~ t}~\epsilon(t) ~+~ 20.53
\end{equation}
with $\epsilon(t)$ as a white noise process with zero mean and variance equal to 1, and 
$20.53$ as the mean value of $LMC_t$ (the mean value of the SDSS $r$-band light curve of 
\obj) (different mean values have few effects on our simulating results), probability of 
long-term light curves of \obj~ related to central AGN activities can be simply discussed. 
With randomly selected $\tau$ from 100days to 1000days (common values for quasars in 
\citealt{mi10}) and collected $\sigma_*$ leading the variance $\tau\sigma_*^2/2$ to be 
0.065 (the variance of $r$-band light curve of \obj) and $t_i$ as the observational time 
shown in top panel of Fig.~\ref{car}, 100000 simulating light curves [$t_i$, $LMC_i$] can 
be created. And, the uncertainties of each $LMC_t$ is simply estimated by 
$LMC_t~\times~LMC_{err}/LMC$ with $LMC$ and $LMC_{err}$ as the r-band light curve and the 
corresponding uncertainties of \obj. Then, among the 100000 artificial light curves, 115 
light curves can be described by the theoretical TDE model, leading to the determined 
$\chi^2/dof$ smaller than 10 (6.5 for the best descriptions to the $r$-band light curve 
of \obj). Therefore, the probability is only about 0.12\% (115/100000)\footnote{Due to 
loose collections of $\tau$, the intrinsic probability should be smaller than 0.12\%.} that 
the detected TDE-expected variability in \obj~ is mis-detected through the intrinsic AGN 
variability. Bottom panel of Fig.~\ref{car} shows one CAR process simulating light curve 
but described by theoretical TDE model. In other words, the TDE expected variability in 
\obj~ are confident enough, at least with confidence level higher than 99.88\% (1-0.12\%) 
(higher than 3$\sigma$). Certainly, jointed the probability through the spectroscopic 
results, the probability is only about $1.3\times10^{-7}$ ($\sim1.1\times10^{-4}\times0.12\%$) 
that the flare properties of \obj~ are related to normal broad line AGN. In other words, 
the confidence level is higher than 5$\sigma$ that the flare properties of \obj~ are not 
related to normal broad line AGN.

	Last but not the least, as recently discussed in \citet{yr23}, the detection rate 
of TDEs is about $R_{TDE}\sim(2-8)\times10^{-5}/galaxy/year$. Here, among the SDSS quasars 
in \citet{sh11}, there are $N=12827$ SDSS quasars covered by the Stripe82 region. And the 
time duration $T_{S82}$ is about 8 years of the SDSS Stripe82 provided light curves as 
described in in \citet{bv08}. Therefore, the expected number $R_{TDE}\times N\times T_{S82}$ 
of TDEs among the SDSS quasars in the Stripe82 region is between 2.1 and 8.2. In this 
manuscript, there are seven TDE candidates, consistent with the expected results by TDEs 
detection rates.

\section{Main Summary and Conclusions}

	The main summary and conclusions are as follows.
\begin{itemize}
\item considering applications of the simple blackbody photosphere model to continuum 
emissions of SDSS quasar, different locations can be confirmed between TDEs and SDSS 
quasars in the parameter space of blackbody temperature $T_{BB}$ versus blackbody radius 
$R_{BB}$. Here, the $T_{BB}$ and the $R_{BB}$ for the TDEs are the values for the peak 
brightness of their light curves.
\item Considering the linear fitting results and 2.6$\sigma$ (99\%) scatters to the $R_{BB}$ 
and $T_{BB}$ of the SDSS quasars, there are 402 SDSS quasars covered by the Stripe82 regions 
accepted as outliers in the parameter space of $R_{BB}$ and $T_{BB}$, which could be treated 
as parent sample of candidates of TDEs.
\item After checking the long-term light curves through the SDSS Stripe82 database, there 
are six objects of which long-term light curves have smooth decline trend, and one object 
(\obj~ at $z=1.16$) of which long-term light curves have steep rise followed by a smooth 
decline trend. The seven objects can be accepted as TDE candidates. 
\item due to the \obj~ having clear information of peak position of the light curves and 
having its SDSS spectrum observed around the peak position for the light curves, it is a 
better chance to check whether the \obj~ is a better TDE candidate. 
\item Through the 7.2-year-long SDSS $ugriz$-band light curves described by theoretical TDE 
model, the TDE model determined $T_{BB}$ and $R_{BB}$ are well consistent with the values 
determined through applications of the blackbody photosphere model to describe its 
spectroscopic continuum emissions around Mg~{\sc ii}, to support a central TDE in \obj.
\item Considering distributions of spectra index and continuum luminosity at 3000\AA~ of 
more than 4900 SDSS quasars around $z\sim1.16$, the probability is only about $10^{-4}$ that 
the spectroscopic continuum emission properties in \obj~ leading to $R_{BB}$ and $T_{BB}$ 
consistent with TDE model determined values are related to continuum emissions of normal quasars. 
\item Considering the properties of CAR/DRW process parameters of $\tau$ and $SF_\infty$ 
for SDSS normal quasars and for the \obj~ of which light curves can be described by the 
JAVELIN code, there are not distinguished locations of the \obj~ from the normal quasars 
in the space of $\tau$ versus $SF_\infty$, indicating it is necessary to check whether the 
long-term variability of \obj~ is actually related to intrinsic AGN variability. 
\item Accepted the CAR process to simulate intrinsic AGN variability properties, confidence 
level higher than $3\sigma$ can be determined to support the SDSS light curves are related 
to a central TDE in \obj.
\item Jointed the simulated results both through the spectroscopic continuum emissions and 
the long-term photometric variability, the confidence level higher than $5\sigma$ can be 
confirmed that the flare properties of \obj~ are related to a central TDE rather than to 
normal broad line AGN.
\end{itemize}

\section*{acknowledgements}
Zhang gratefully acknowledge the anonymous referee for giving us constructive 
comments and suggestions to greatly improve our paper. Zhang gratefully acknowledges the 
kind financial support from GuangXi University, and the grant support from NSFC-12173020 
and NSFC-12373014. This manuscript has made use of the data from the SDSS projects, managed 
by the Astrophysical Research Consortium for the Participating Institutions of the SDSS-III 
Collaboration. The manuscript has made use of the public code of TDEFIT 
(\url{https://github.com/guillochon/tdefit}) and MOSFIT 
(\url{https://github.com/guillochon/mosfit}), and use of the MPFIT package 
(\url{http://cow.physics.wisc.edu/~craigm/idl/idl.html}) written by Craig B. Markwardt, and 
use of the LtsFit package 
(\url{https://www-astro.physics.ox.ac.uk/~cappellari/software/#sec:lts}) written by Michele 
Cappellari. The manuscript has made use of the data from the ZTF 
(\url{https://www.ztf.caltech.edu/}).


\label{lastpage}
\end{document}